\documentclass{sf2a-conf2012}
\usepackage{graphicx}
\usepackage{hyperref}
\usepackage[]{natbib}  
\usepackage[cyr]{aeguill}
\usepackage{epstopdf}

\def\BibTeX{{\rm B\kern-.05em{\sc i\kern-.025em b}\kern-.08em
    T\kern-.1667em\lower.7ex\hbox{E}\kern-.125emX}}
\bibpunct{(}{)}{;}{a}{}{,}  %%%%%%%%%%%%%  A&A bibliography style
\begin{document}

\TitreGlobal{SF2A 2012}

%%-----------------------------------------------------------------
%%      the top matter
%%

\title{Stars with the B[e] phenomenon seen by long baseline interferometry}

\runningtitle{Stars with the B[e] phenomenon}

\author{M. Borges Fernandes}\address{Observat\'orio Nacional/MCTI, Rio de Janeiro, Brazil}

\author{O. Chesneau}\address{Laboratoire Lagrange, Universit\'e de Nice Sophia-Antipolis, CNRS, Observatoire de la C\^ote d'Azur, France}

\author{M. Kraus}\address{Astronomick\'y \'ustav, Ond\v{r}ejov, Czech Republic}

\author{L. Cidale}\address{Universidad Nacional de La Plata, Argentina}

\author{A. Meilland$^2$}

\author{P. Bendjoya$^2$}

\author{A. Domiciano de Souza$^2$}

\author{G. Niccolini$^2$}

\author{I. Andruchow$^4$}

\author{S. Kanaan}\address{Universidad de Valpara\'iso, Chile}

\author{P. Stee$^2$}

\author{F. Millour$^2$}

\author{A. Spang$^2$}

\author{M. Cur\'e$^5$}

%% Keep this line, even if the page will be settled afterwards.
\setcounter{page}{237}
\maketitle

%%-----------------------------------------------------------------
%%        The abstract
%% 
%%  Warning!  within the abstract:
%%  - do not use macros. 
%%  - do not use commands like: \cite, \citet, \citep ... etc.

\begin{abstract}
Thanks to the high spatial resolution provided by long baseline interferometry, it is possible to understand the complex circumstellar geometry around stars with the B[e] phenomenon. These stars are composed by objects in different evolutionary stages, like high- and low-mass evolved stars, intermediate-mass pre-main sequence stars and symbiotic objects. However, up to now more than 50\% of the confirmed B[e] stars are not well classified, being called unclassified B[e] stars. From instruments like VLTI/AMBER and VLTI/MIDI, we have identified the presence of gaseous and dusty circumstellar disks, which have provided us with some hints related to the nature of these objects. Thus, we show our results for two galactic stars with the B[e] phenomenon, HD\,50138 and CPD\,52$^o$9243, based on interferometric measurements. 
\end{abstract}

%% Insert the keywords (to appear in the ADS indexing)
%% Keywords must be separated by a comma

%%-----------------------------------------------------------------

\section{Introduction}
%%---------------------
The B[e] phenomenon was defined by the presence in the optical spectra of B-type stars \citep{Conti1997} of strong Balmer emission lines, and permitted and forbidden emission lines of neutral and singly ionized metals, like Fe\,II and O\,I. In addition, these stars also present a strong near or mid-infrared excess due to hot circumstellar dust. Based on \citet{Lamers1998}, there are different types of objects presenting the B[e] phenomenon: pre-main sequence Herbig Ae/Be stars, compact planetary nebula, symbiotic objects, hot supergiants, and unclassified objects, whose evolutionary stage is still unknown.

Recently the optical/IR long baseline interferometry, especially with the VLTI, has become an important tool to deeply study the circumstellar environment of the brightest B[e] stars. Thanks to the field of view and the high spatial resolution of instruments, like AMBER and MIDI, it is possible to have access to information related to circumstellar medium close to the stars, allowing us to obtain sizes, shapes, and orientations, as a function of wavelength in the optical, near and mid-IR ranges. 

In this paper, we present our results for two galactic stars with the B[e] phenomenon, namely HD\,50138 and CPD\,52$^o$9243, based on interferometric measurements.

\section{Results}
%%-------------------------

\subsection{HD\,50138}
%%---------------------------------
 
HD 50138 is one of the closest and brightest B[e] stars already identified in our galaxy. However its evolutionary stage is still not well known, being classified in the literature as either a classical Be or a pre main-sequence Herbig star. This object presents strong spectral variations that seem to be associated to outbursts and shell-phases. From a high-resolution spectroscopic analysis (FEROS and Narval data) and using photometric data from the literature, it was found that a new shell-phase happened before 2007 \citep{BF2009}. In addition, spectro-polarimetric data from the literature have indicated the presence of a non-spherically symmetric circumstellar environment: probably a disk.   

We obtained data from 14 baselines using MIDI, 17 using AMBER and also from 92 measurements at 10.7 $\mu$m from the KECK segment-tilting experiment \citep{Monnier2009}. We modeled these data using analytical geometrical models and the best results were found assuming a similar geometry for both near and mid-IR data: two elliptical Gaussian distributions (see Figs.\,1 and 2 and Table\,1).

\begin{figure}[ht!]
 \centering
 \includegraphics[width=0.8\textwidth,clip]{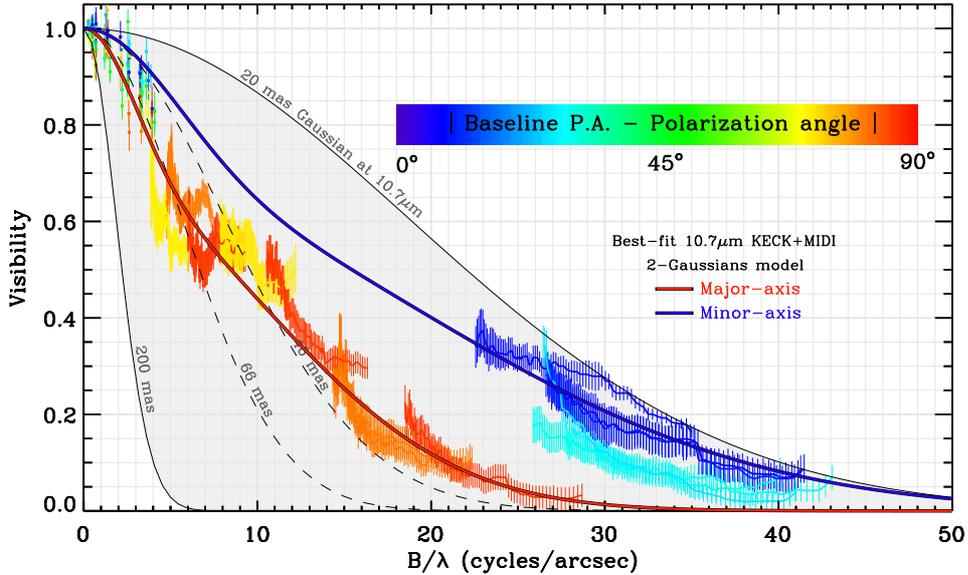}      
%% Note the ABSENCE of the extension .pdf , .eps or .ps  !
  \caption{VLTI/MIDI visibilities plotted as a function of the spatial frequency, B/$\lambda$, with kind permission from \citet{BF2011}. The colors indicate the baselines position angle with respect to the polarization angle derived by \citet{YE1998}. The color scale (i.e., blue, cyan, green, yellow, orange, and red) goes from blue for baselines parallel to the polarization angle to red for the perpendicular ones. The dots with error bars indicate the Keck measurements. The thin solid and dashed lines represent the visibilities from Gaussian disks with different FWHM. The thick red solid line represents the model fitting aligned to the PA of the major axes of the 2-Gaussians distribution and the blue one to the PA of the minor axes. }
  \label{fig1}
\end{figure}

\begin{table}[!htb]
\caption{Parameters of the best-fit models from our analysis of interferometric data from HD\,50138, assuming two elliptical Gaussian distributions, where ``G1" means the first Gaussian distribution, ``G2" is second one, ``i" means inclination, and ``$\theta$" is the projection of the major axis onto the sky plane, see \citet{BF2011}.}
{\centering \begin{tabular}{cccc}
\hline
  & MIDI+KECK & AMBER \\ 
\hline
Flux$_{star}$  &     &  0.12 $\pm$ 0.01 \\
Flux$_{G1}$ & 0.68 $\pm$ 0.04 & 0.61 $\pm$ 0.06 \\
FWHM$_{major-G1}$ (mas) & 35.2 $\pm$ 1.5 & \\
FWHM$_{major-G2}$ (mas) & 131.4 $\pm$ 11.2 & \\
FWHM$_{minor-G1}$ (mas) &   & 3.0 $\pm$ 0.4 \\
FWHM$_{minor-G2}$ (mas) &  & $\ge$ 14.0 \\
Flattening   &   1.82 $\pm$ 0.02 & 1.7 $\pm$ 0.3 \\
$\theta$ ($^o$)& 65.9 $\pm$ 2.0 & 77 $\pm$ 2 \\
$i$ ($^o$) & 56.7 $\pm$ 0.4 & 54 $\pm$ 8 \\
\hline
$\chi^{2}_{r}$ & 1.9 & 13.3 \\
\hline
\end{tabular}\par}
\label{models} 
\end{table}

\begin{figure}[ht!]
\centering\includegraphics[width=0.8\textwidth]{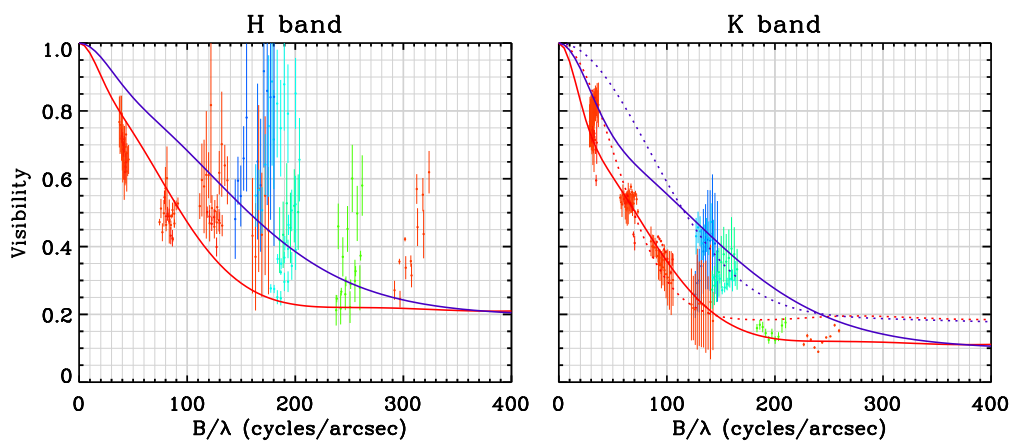}
\caption{Model fitting of the VLTI/AMBER data of HD\,50138, with kind permission from \citet{BF2011}, obtained using LITpro software (available at \url{http://www.jmmc.fr/litpro}): right panel for the K-band assuming two elliptical Gaussian distributions. The different colors follow the same description as in Fig.\,1, representing the orientation of the baselines. } 
\label{amberKmodel2}
\end{figure}

Based on our results, the presence of a disk composed by gas and dust, with an inclination of $\sim$ 56$^o$ and a projection on the sky plane of $\sim$ 71$^o$ was confirmed \citep{BF2011}.

The formation mechanism of this disk is not clear yet, but it could be linked, as for classical Be stars, to stellar pulsations \citep{BF2012}.

\subsection{CPD\,52$^o$9243}
%%-------------------

CPD-52$^o$9243 is an emission-line star, which has been classified as a B[e] supergiant candidate. However, its stellar parameters are still not well known. Recently we obtained MIDI data from 4 baselines using the UTs and from an analytical geometrical modeling, we could find evidence for the presence of a dusty disk around this object \citep{Cidale2012}.

The disk has an inclination angle of 46$\pm$7 degrees, and an upper limit of its dusty inner edge of $\sim$ 8\,mas (see Fig.\,3 and Table\,2). Furthermore, the dusty disk surrounds a cool, detached molecular ring seen in CO band emission. The scenario for CPD-52$^o$9243 might therefore be quite similar to other B[e] supergiants, identified in close binaries, and found to have circumbinary molecular and dusty disks: HD 327083 \citep{Wheelwright2012} and GG\,Car \citep{Kraus2012}.

\begin{figure}[!ht]
\centering   
\includegraphics[width=0.4\textwidth,clip]{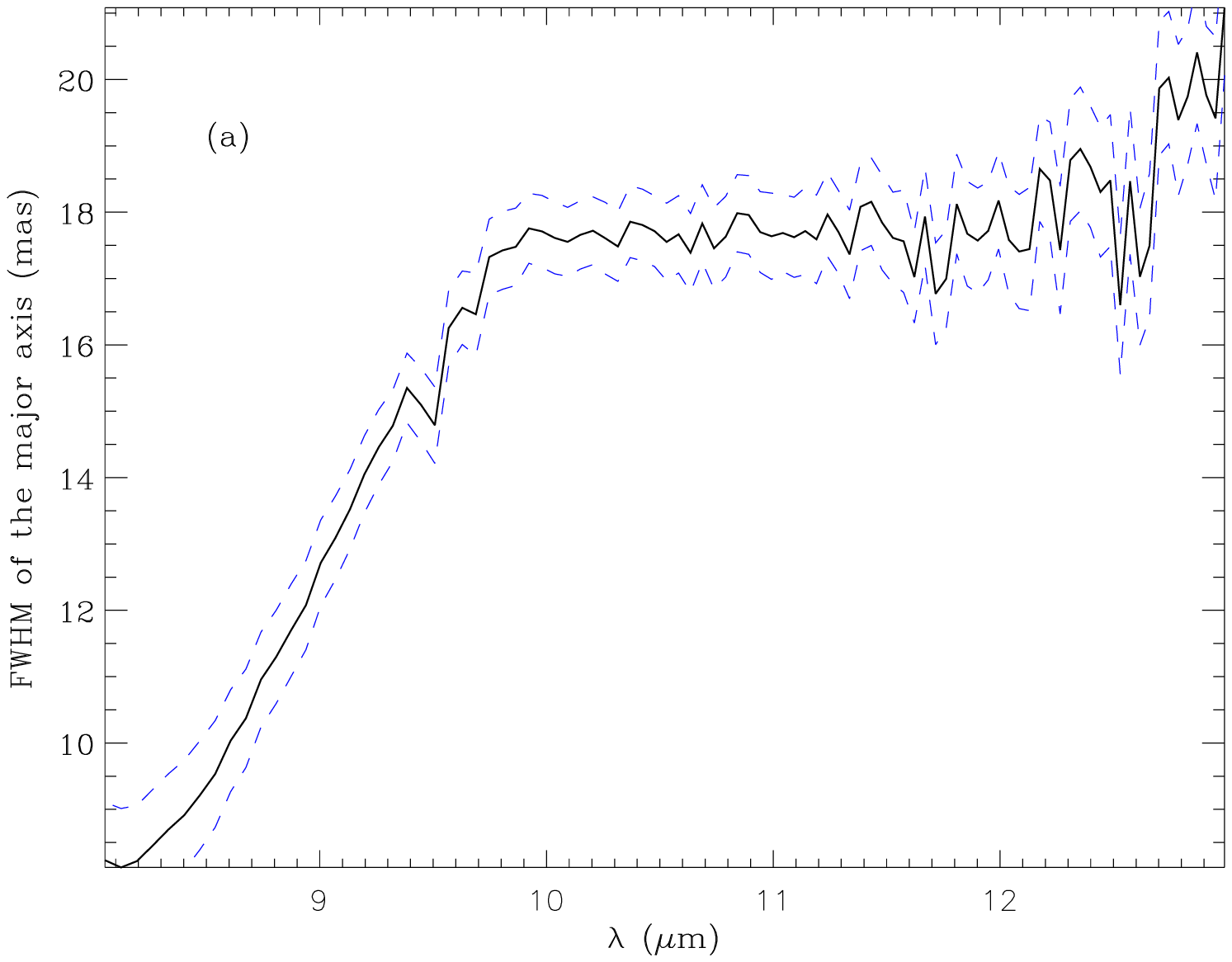}
\includegraphics[width=0.4\textwidth,clip]{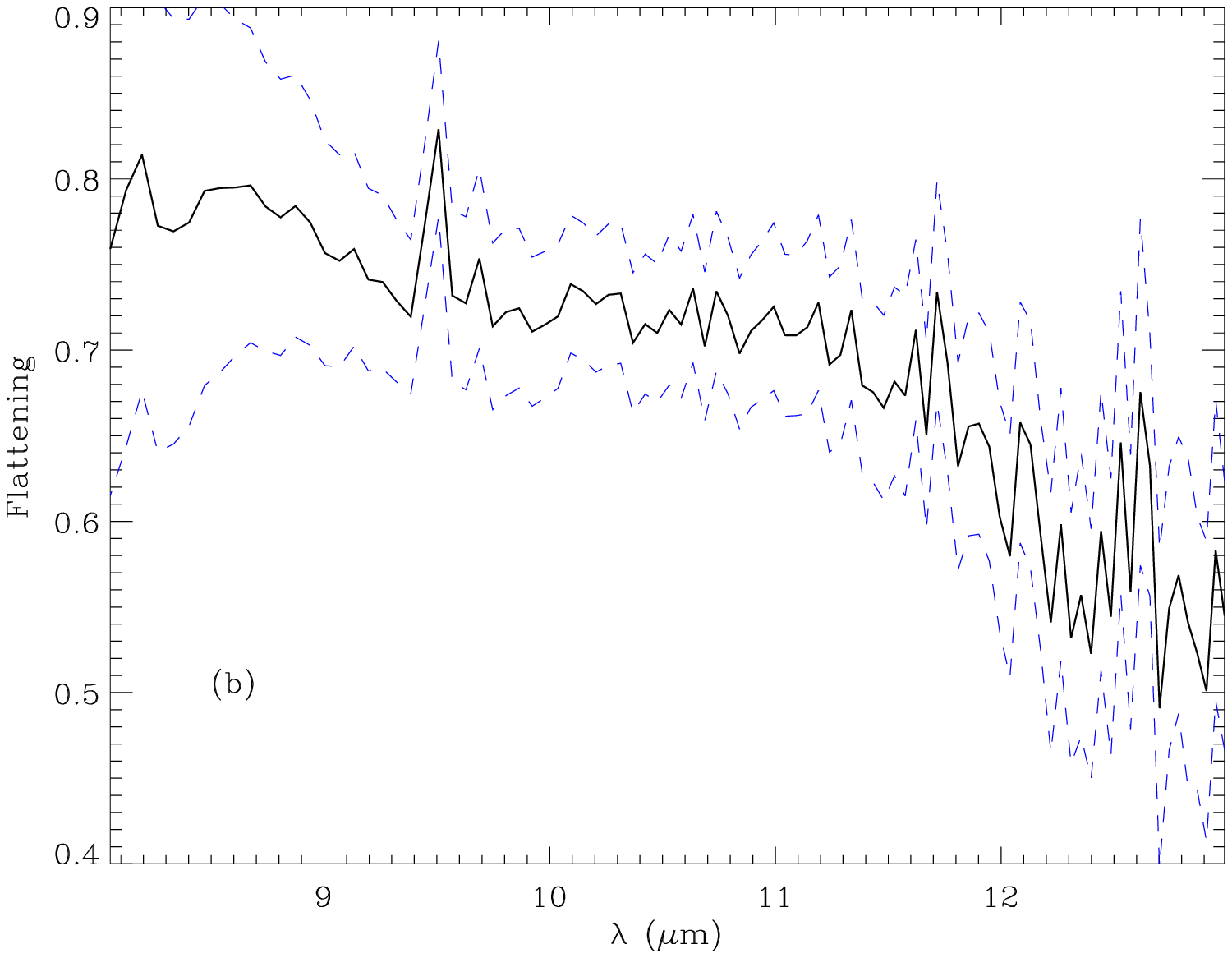}
\includegraphics[width=0.4\textwidth,clip]{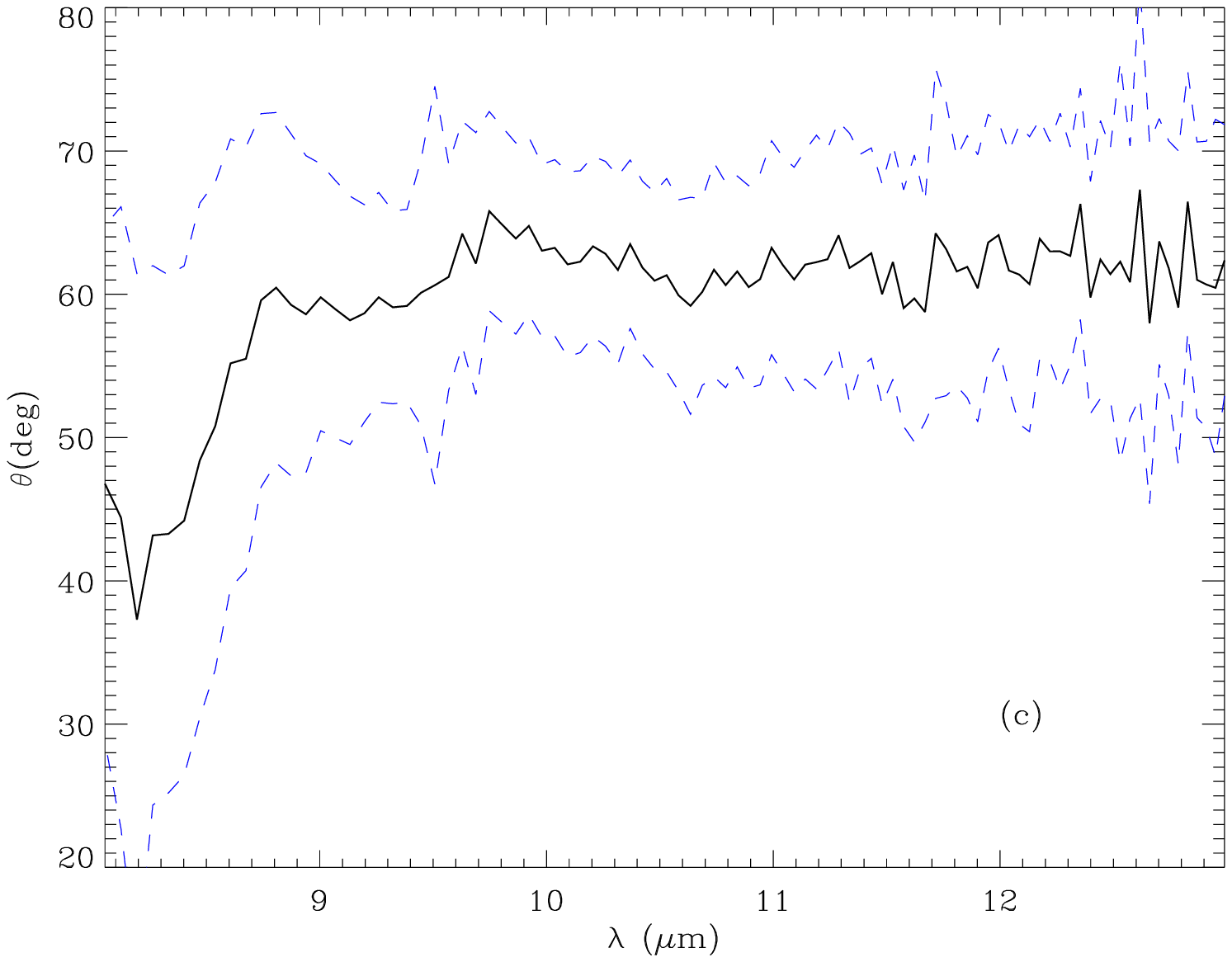}
\includegraphics[width=0.4\textwidth,clip]{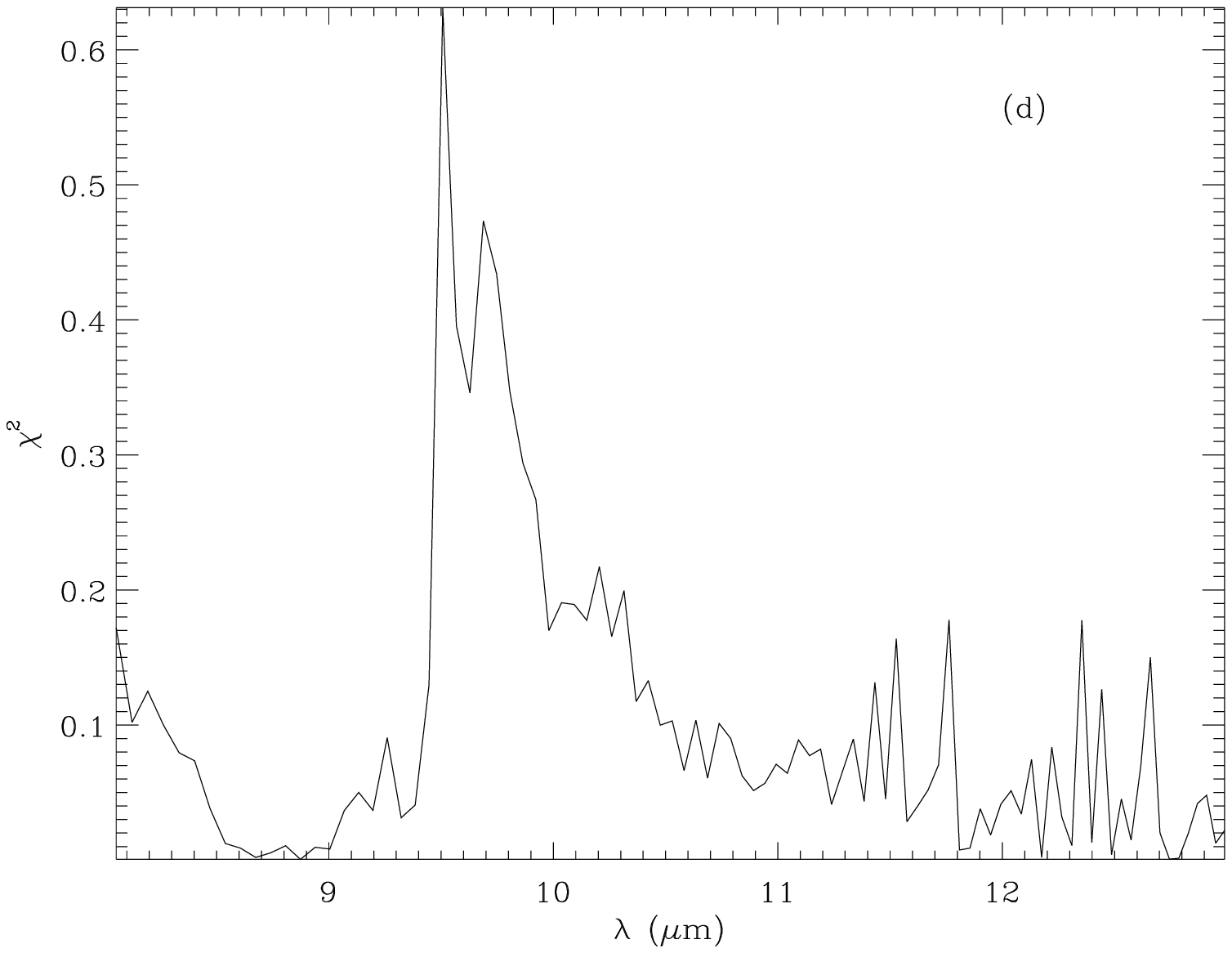}
\caption{The parameters provided by the best model, considering an elliptical Gaussian distribution for the circumstellar medium of CPD-52$^o$9243, which fits the VLTI/MIDI baselines, as a function of the wavelength: (a) the FWHM of the major axis; (b) the flattening (ratio of FWHM of the minor axis and of the major axis); (c) the projection of the major axis ($\theta$) onto the sky plane; and (d) the reduced $\chi^2$ for each wavelength. The dashed lines indicate the errors for each parameter, as a function of the wavelength.}
\label{model2}
\end{figure}

\begin{table}[!htb]
\caption{Parameters of our best-fitting model for VLTI/MIDI data of CPD-52$^o$9243.``$\theta$" is the projection of the major axis onto the sky plane.}
{\centering \begin{tabular}{cc}
\hline\hline
FWHM major axis (mas) & 16.35 $\pm$ 0.73 \\
Flattening & 0.69 $\pm$ 0.07 \\
$\theta$ ($^o$) & 60.13 $\pm$ 9.57 \\
$\chi^2$ & 0.10 \\
\hline 
\end{tabular}\par}
\label{Tabla_CPD2}
\end{table}

\section{Conclusions}
%%--------------------
Optical long baseline interferometry is an ideal tool to provide information concerning the geometry and inclination of the circumstellar matter of stars with the B[e] phenomenon. These results will certainly contribute to a better comprehension of the nature of some unclassified B[e] stars and the inclusion of the B[e] phenomenon in the evolutionary tracks.

% Optional acknowledgements
% -------------------------
\begin{acknowledgements}
MBF acknowledges Conselho Nacional de Desenvolvimento Cient\'{\i}fico e Tecnol\'ogico (CNPq-Brazil) for the post-doctoral grant. LC acknowledges financial support from the Agencia de Promoci\'on Cient\'{\i}fica y Tecnol\'ogica (BID 1728 OC/AR PICT 111), from CONICET (PIP 0300), and the Programa de Incentivos G11/089 of the Universidad Nacional de La Plata, Argentina. MK acknowledges financial support from GA\,\v{C}R under grant number 209/11/1198. The Astronomical Institute of Ond\v{r}ejov is supported by the project RVO:67985815. I.A. acknowledges financial support from a posdoctoral fellow of CONICET. MC acknowledges financial support from Centro de Astrof\'{\i}sica de Valpara\'{\i}so.
\end{acknowledgements}

%%-----------------------------
%%   Bibliography
%%-----------------------------
%%
%% The reference list should contain all the references cited in the text, ordered alphabetically by surname (with
%% initials following). If there are several references to the same first author, they should be entered according
%% to the following scheme:
%% 1. One author: chronologically
%% 2. Author, one co-author: alphabetically by co-author, then chronologically
%% 3. Author, two or more co-authors: chronologically.
%%
%% Please note that for papers that have more than five authors, only the first three should be given, followed
%% by "et al."
%%
%% The format for references is the one adopted by A&A (see the example below).
%%
%% To set the reference list in the proper A&A format, we encourage you to use BibTEX and the natbib
%% package instead of the standard 'thebibliography' environment.
%%

%% The following lines are required when using BibTEX (strongly encouraged!):
%\bibliographystyle{/data/conferencias/internacionais/SF2A-2012/proceedings/aa}  % A&A bibliography style file (aa.bst)
%\bibliography{sf2a} % your references in file: Yourfile.bib

%
\end{document}